\documentclass[aps,prb,showpacs,twocolumn,superscriptaddress]{revtex4-1}
\usepackage{amsmath}
\usepackage{amssymb}
\usepackage{graphicx}
\usepackage{hyperref}
\usepackage{url}
\begin{document}
\title{Quantum-Confinement-Induced Magnetism in LaNiO$_3$-LaMnO$_3$ Superlattices}
\author{Shuai Dong}
\affiliation{Department of Physics and Astronomy, University of Tennessee, Knoxville, Tennessee 37996, USA}
\affiliation{Materials Science and Technology Division, Oak Ridge National Laboratory, Oak Ridge, Tennessee 37831, USA}
\affiliation{Department of Physics, Southeast University, Nanjing 211189, China}
\author{Elbio Dagotto}
\affiliation{Department of Physics and Astronomy, University of Tennessee, Knoxville, Tennessee 37996, USA}
\affiliation{Materials Science and Technology Division, Oak Ridge National Laboratory, Oak Ridge, Tennessee 37831, USA}
\date{\today}

\begin{abstract}
The emergence of magnetic reconstructions
at the interfaces of oxide heterostructures
are often explained via subtle modifications in the
electronic densities, exchange couplings, or strain.
Here an additional possible route for induced magnetism is studied
in the context of the (LaNiO$_3$)$_n$/(LaMnO$_3$)$_n$ superlattices using a hybrid tight-binding model.
In the LaNiO$_3$ region,
the induced magnetizations decouple from the intensity
of charge leakage from Mn to Ni,
but originate from the spin-filtered quantum confinement present in these
nanostructures. In general, the induced magnetization is the largest for the
(111)-stacking and the weakest for the (001)-stacking superlattices,
results compatible with the exchange bias effects reported by
Gibert \textit{et al}. Nat. Mater. \textbf{11}, 195 (2012).
\end{abstract}
\maketitle

\section{Introduction}
The area of research that focuses on oxide heterostructures is attracting considerable attention because of its importance in the development of quantum devices based on correlated electronic systems.\cite{Mannhart:Sci,Takagi:Sci,Hammerl:Sci,Dagotto:Sci07}
Novel physical properties are expected to emerge from the electronic reconstruction near the interfaces.\cite{Bibes:Ap,Martin:Mse,Zubko:Arcmp,Hwang:Nm}
In particular, the interfacial magnetism can have properties different from those of
bulk materials and in recent years several investigations revealed various magnetic
reconstructions. Their origin can be mainly classified via mechanisms involving
modifications in the (1) electronic densities, (2) exchange couplings, and/or (3) strain.
For example, mainly due to strain and charge transfer, pure manganite LaMnO$_3$/SrMnO$_3$ (LMO/SMO)
superlattices (SLs) display a variety of magnetic
orders,\cite{Bhattacharya:Prl,May:Nm,May:Prb,Dong:Prb08.3,Zhang:Prb12,Dong:Prb12}
while due to modifications in the exchange coupling, distinct magnetic states
emerge in the (001)- (011)- and (111)-stacking of LaFeO$_3$/LaCrO$_3$
SLs.\cite{Ueda:Sci,Ueda:Jap,Zhu:Jap} Furthermore, the interfacial
magnetic orders of La$_{1-x}$Sr$_{x}$MnO$_3$ (LSMO) can be tuned by attaching
ferroelectric layers (e.g. LSMO/BiFeO$_3$, LSMO/BaTiO$_3$, and LSMO/PZT),
that induce modifications in the interfacial electronic density.\cite{Yu:Prl,Vaz:Prl,Burton:Prb,Burton:Prl,Calderon:Prb11,Dong:Prb11}

\begin{figure}
\includegraphics[width=0.47\textwidth]{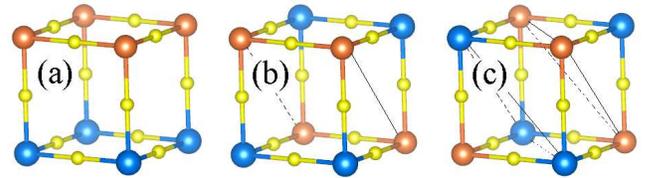}
\caption{(Color online) The three stacking configurations of perovskite SLs (period $n=1$)
studied here: (a) (001); (b) (011); (c) (111). Blue and brown balls
are the transition metals, such as Ni and Mn, while the yellow balls
are the oxygens. The A-site cation, e.g. La, is not shown.}
\label{crystal}
\end{figure}

Recently, an exchange bias effect was reported
experimentally in (LaNiO$_3$)$_n$/(LaMnO$_3$)$_n$ (LNO/LMO)
SLs grown along the (111) axis
of a pseudocubic structure.\cite{Gibert:Nm} In contrast,
no exchange bias was observed in the conventional
(001)-stacking of LNO/LMO SLs,\cite{Gibert:Nm} suggesting a qualitative difference
between the (001)- and (111)-stacking directions despite having the same compositions and periodicity (as sketched in Fig.~\ref{crystal}). This exchange bias
in the (111)-stacking is nontrivial
since it is well known that LNO
 is a paramagnetic (PM) metal
in its bulk form.\cite{Catalan:Pt} Therefore, interesting physical questions arise:
what is the origin of the induced magnetic moments in the LNO layers
of these SLs? Are they in proportion to the charge transferred across
the interfaces, considering that electrons in the LMO layers are spin polarized?
Why are the (001)- and (111)-SLs
qualitatively different with regards to their magnetic properties?
And what are the expected results for other stacking orientations not yet explored,
such as the (011)-direction?

In this manuscript, LNO/LMO SLs will be studied theoretically from the
perspective of microscopic models. Our main result is that the experimentally
observed magnetism in the LNO layers
appears to be mainly caused by \emph{quantum-confinement} effects.
This induced magnetism is weakly coupled with
the actual value of the charge that is transferred from Mn to Ni, and depends
strongly and nonlinearly on the stacking
orientations and the SL periodicity.
The underlying physical mechanism discussed here
can partially explain the nontrivial exchange bias observed
in (LNO)$_n$/(LMO)$_n$ SLs.

\section{Model \& Method}
A hybrid two-orbital tight-binding model
is here constructed, containing the Hubbard interaction for both the manganite
and nickelate components and the double-exchange (DE) term for the manganite
sector only. In past decades,  extensive investigations have shown that
the two-orbital DE model is a successful model to describe manganites,\cite{Dagotto:Prp,Dagotto:Njp}
while the two-orbital Hubbard model has also been often employed
for the nickelates.\cite{Hotta:Prl04,Lee:Prb} Recent theoretical studies also used these
two models for manganite heterostructures and for LNO bilayers.\cite{Dong:Prb08.3,Yu:Prb09,Yang:Prb,Ruegg:Prb,Ruegg:Prb12}
For these reasons, our model provides a reasonable
starting point to address the LNO/LMO SLs.
More explicitly, the model Hamiltonian used here can be written as:
\begin{eqnarray}
\label{hami}
\nonumber
H&=&\sum_{<ij>,\sigma}^{\alpha\beta}t^{\vec{r}}_{\alpha\beta}(c_{i\alpha\sigma}^{\dagger}c_{j\beta\sigma}+H.c.)
-\frac{J_{\rm H}}{2}\sum_{i\in\rm Mn}n_i\vec{\sigma}_i\cdot\vec{S}_i\\
&&+\sum_{i\in{\rm Ni}}V_{\rm Ni}n_i+H_{\rm Hubbard}(U,J),
\end{eqnarray}
where the first term is the standard nearest-neighbor hopping (i.e. the kinetic energy) between orbital $\alpha$
of site $i$ and orbital $\beta$ of site $j$. Here the $e_{\rm g}$
orbitals $d_{x^2-y^2}$ (=1) and $d_{3z^2-r^2}$ (=2) are
employed since in both LaMnO$_3$ and LaNiO$_3$ the transition metals are
in the $e_{\rm g}^1$ configuration. The (standard) hopping amplitudes are
orbital and direction dependent: $t^x_{11}=t^y_{11}=3t^x_{22}=3t^y_{22}=-3t_0/4$;
$t^x_{12}=t^x_{21}=-t^y_{12}=-t^y_{21}=\sqrt{3}t_0/4$; $t^z_{11}=t^z_{12}=t^z_{21}=0$; $t^z_{22}=-t_0$.
In the  present work, the hopping unit $t_0$ is assumed to be the same
for the bonds Mn-O-Mn, Ni-O-Ni, as well as Mn-O-Ni.
This approximation is reasonable since density functional theory (DFT)
studies of LaNiO$_3$ led to\cite{Ruegg:Prb12} $t_0 \sim 0.6$~eV while a
very close result $t_0 \sim 0.5\sim0.6$~eV was found for
LaMnO$_3$.\cite{Franchini:Jpcm} Moreover,
$t_0$ will be taken as the energy unit in this work.

The Hund's coupling second term affects only the Mn ions that are in a
high-spin $t_{\rm 2g}^3$ configuration. Then the spin-up and -down levels are split by $J_{\rm H}$ in LaMnO$_3$.
Here, $J_{\rm H}$ is set to be $4t_0$ ($\sim2-2.4$ eV)\cite{Dagotto:Prp}
which is large enough to induce half-metal behavior in manganites.
In contrast, the Ni ions are in the $t_{\rm 2g}^6$ configuration.
Then the spin-up and -down levels are degenerate in LaNiO$_3$, suggesting a non-magnetic background.

The third term is the on-site $e_{\rm g}$ potential difference between Ni and Mn.
Considering the potential of the Mn's spin-up level as the zero of reference,
the potential of the Mn's spin-down level becomes $J_{\rm H}$ due to
the Hund coupling, while the potential $V_{\rm Ni}$ for the Ni levels
will be varied as a parameter in our investigations.

The last term is the standard Hubbard interaction
for multi-orbital models acting in the whole lattice, with the parameters $U$
(intra-orbital Coulomb repulsion) and $J$ (inter-orbital Hund exchange).\cite{Hotta:Prl04,Lee:Prb,Yang:Prb,Ruegg:Prb,Ruegg:Prb12} This term is
here treated using the Hartree mean-field
approximation, which is a quite reasonable starting point to handle Hubbard interactions
in these systems according to previous literature.\cite{Yang:Prb,Ruegg:Prb,Ruegg:Prb12}
The widely used ratio $J=U/4$ is used,
and $U$ is tuned as a parameter for both the manganite and nickelate layers.
In reality, the value of $U$ may be different between Mn and Ni.
However, the physical consequences of $U$ are not important
 for the $e_{\rm g}$ sector of manganites: at least in the
Hartree mean-field approximation it has been shown that  a large $J_{\rm H}$ coupling
already plays a similar role.\cite{Dagotto:Bok} Then, to avoid a large number of
tunable parameters, here the same $U$ is applied to both the
manganite and nickelate layers. LaNiO$_3$ is a PM metal,
implying a weak $U$ (otherwise it would become magnetically ordered).\cite{Gibert:Nm}
Then, in the present work $U$ will be tuned between $0$ and $2t_0$. Note that larger
values of $U$, such as $3t_0$, have also been tested and the results do not alter
our physical conclusions qualitatively.

Partially due to strain, the ground state of LMO ultra-thin films
with nearly cubic structure grown on SrTiO$_3$
are ferromagnetic (FM).\cite{Gibert:Nm,Bhattacharya:Prl,Dong:Prb08.3}
In addition, $e_{\rm g}$ electrons leak from Mn to Ni,
altering the Mn valence to $+(3+\delta)$, further
driving the LMO layers to a FM state
according to the phase diagram of manganites.\cite{Dagotto:Prp}
Thus, the $t_{\rm 2g}$ spin background of the Mn layers
is set here as FM unless explicitly noted.
For this first study, lattice distortions are neglected, assumption also used
in other previous theoretical studies of LaNiO$_3$-based
heterostructures.\cite{Yang:Prb,Ruegg:Prb,Ruegg:Prb12}
Then, Eq.~\ref{hami} is solved
self-consistently at zero temperature on $4\times4\times L$ lattices with twisted boundary conditions,\cite{Dong:Prb11,Dong:Prb12}
where $L$ is determined by the period of each SL.

\section{Results \& Discussion}
\subsection{Non-correlated limit}
\begin{figure}
\includegraphics[width=0.35\textwidth]{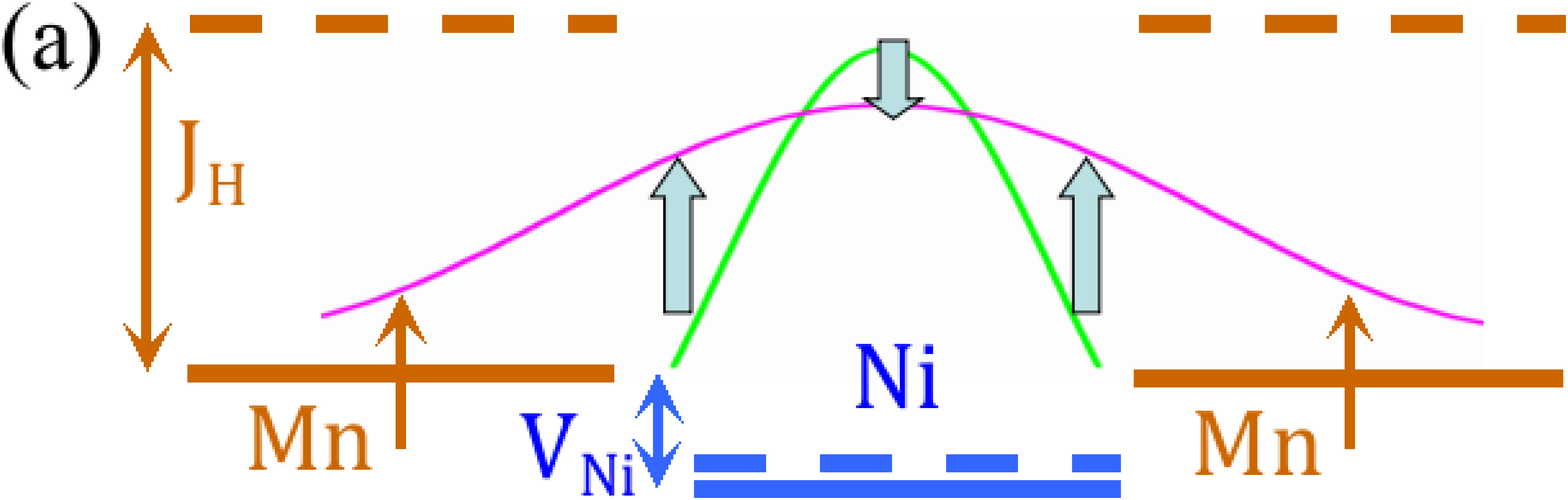}
\includegraphics[width=0.5\textwidth]{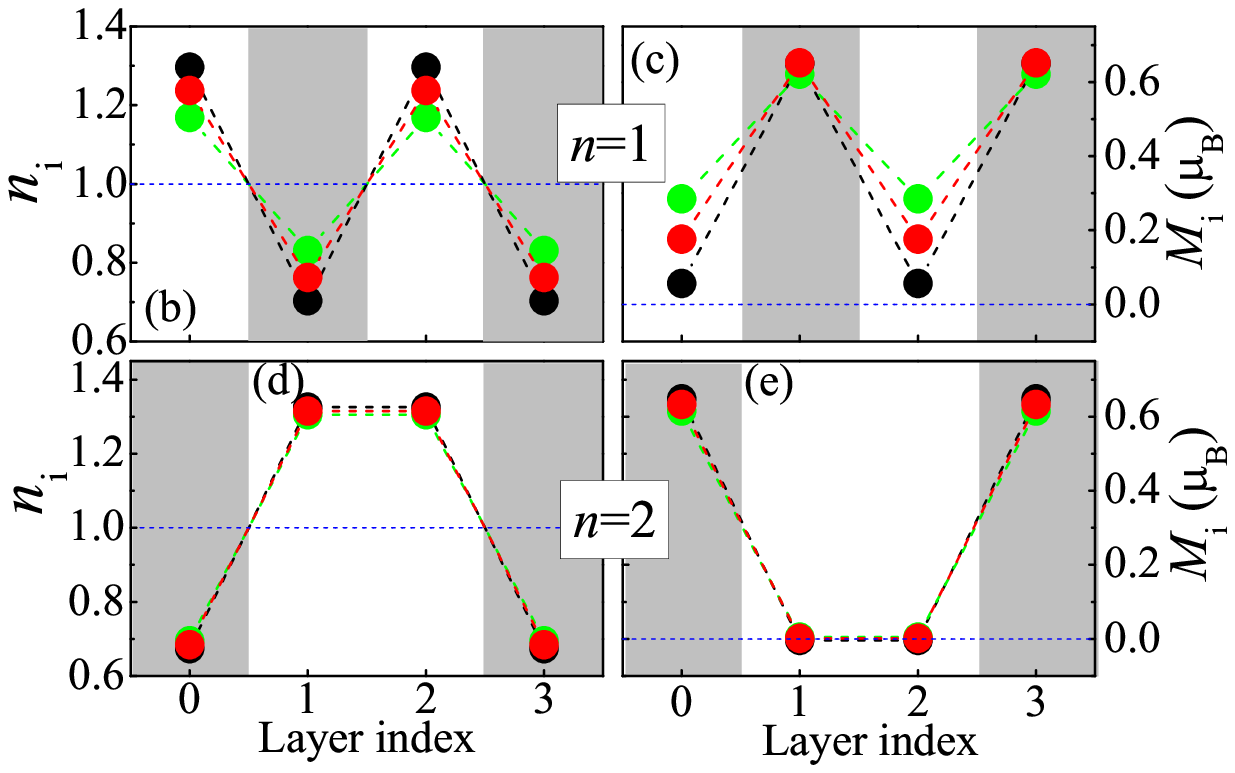}
\caption{(Color online) (a) Sketch of spin-filtered quantum confinement.
Solid and broken lines denote the spin-up and spin-down levels,
which are degenerate for Ni but split by $J_{\rm H}$ for Mn. $V_{\rm Ni}<0$ implies lower Ni levels. Pink and green curves are the sketch of spin-up and spin-down $e_{\rm g}$ electron densities. (b-e) The $e_{\rm g}$-density profiles (left) and $e_{\rm g}$ magnetization (right) of SLs at $V_{\rm Ni}=U=0$. Black: (001)-; Red: (011)-; Green: (111)-stacking SLs. Gray: LMO; White: LNO.}
\label{nm}
\end{figure}

First, the simplest case ($U=0$, $V_{\rm Ni}=0$) is studied.
This non-correlated limit ($U=0$) is useful to clarify the underlying physics,
and $V_{\rm Ni}=0$ means that there is no confinement for the spin-up channel
although the spin-down channel is still confined due to the large Hund $J_{\rm H}$ barrier,
as sketched in Fig.~\ref{nm}(a). To start the discussion,
let us focus first on short periodic ($n\leqslant2$) SLs
because it will be shown below that the induced magnetism is uniform in this case.

Previous DFT calculations\cite{Gibert:Nm} indicated that
at the interface the $e_{\rm g}$ electrons transfer
from Mn$^{3+}$ to Ni$^{3+}$.
As shown in Fig.~\ref{nm}(b) and (d), our calculations are in agreement
since even at $V_{\rm Ni}=0$, $e_{\rm g}$ electrons do
leak from Mn$^{3+}$ to Ni$^{3+}$. This is because the $e_{\rm g}$ levels in Mn's sites are spin polarized,
pushing the Fermi level of LMO to higher energies since only spin-up bands can be occupied
while spin-down bands are almost empty. By contrast,
in LNO both the spin-up and -down bands can be filled,
accommodating more electrons with a lower Fermi level.

The charge transfer from LMO to LNO depends on both the period $n$ as well
as the stacking orientations. For the $n=1$ case (Fig.~\ref{crystal}),
the (001)/(011)/(111)-stacking SLs
have two/four/six Ni-Mn but four/two/zero Mn-Mn or Ni-Ni nearest-neighbors per site.
For other cases ($n\geqslant2$), each interfacial Mn (Ni) ion has one/two/three Ni (Mn)
nearest-neighbors but five/four/three Mn (Ni)
nearest-neighbors in the (001)/(011)/(111)-stacking SLs.
Therefore, naively, the charge transfer
from Mn to Ni may be the strongest (weakest)
in the SLs with (111)-stacking [(001)-stacking]
since they have the most (least) Mn-Ni bonds.

However,
this naive scenario is too simplistic. For example, for $n=2$ the charge transferred to LNO
is nonzero, and almost identical, for the three orientations. But at the same time
the associated LNO magnetization is nearly vanishing. Moreover,
in the $n=1$ case the charge transferred to the Ni-layers is the highest for
the (001)-stacking, yet the magnetization is the {\it smallest} for the same stacking.
Therefore, the intensity of charge leakage is not in linear proportion
to the induced magnetization.

According to Fig.~\ref{nm}, there are several
interesting features in the induced magnetism of
LNO in these short-period SLs. In the $n=1$ SLs, the (111)-stacking shows
the most prominent induced magnetization, while the (001)-stacking is nearly
non-magnetic and the (011)-stacking interpolates between (001) and (111).
In the $n=2$ cases, the induced magnetism of Ni is almost zero
irrespective of the stacking orientations

\begin{figure}
\includegraphics[width=0.55\textwidth]{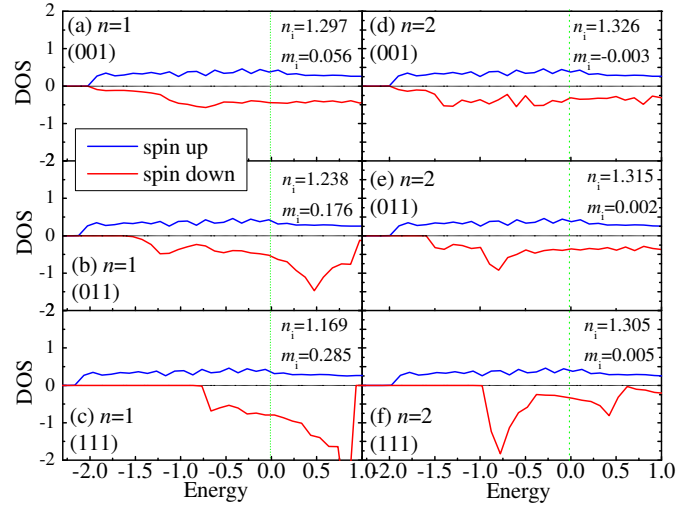}
\caption{(Color online) (a-f) Spin-resolved density of states of Ni. The Fermi level is at $0$. Here $U=0$ and $V_{\rm Ni}=0$. Left: $n=1$; Right: $n=2$. Upper: (001)-stacking; Middle: (011)-stacking; Lower: (111)-stacking. The $e_{\rm g}$ electron densities and magnetism are also shown.}
\label{dos}
\end{figure}

These features also imply that the induced magnetism is indeed \emph{not} simply directly
correlated with the leakage of spin-polarized charge. For
a better understanding of our
results consider instead the band structures of our SLs.
As shown in Fig.~\ref{dos}, the spin-resolved density of states (DOS) at the Ni layers
shows that the spin-up and -down channels are notoriously different.
Due to the high spin polarization of manganites, the spin-up $e_{\rm g}$ electrons
spread much farther in the SLs than the spin-down $e_{\rm g}$ electrons, which
are mostly confined
to the LNO layers. From this perspective, the LMO layers act
as atomic-scale spin filters, causing the local band structures of the LNO layers
to be quite different between the spin-up and -down channels.
Furthermore, this quantum confinement severely depends on the stacking
orientations and periodicity.
The confinement is the most effective
(i.e. with the narrowest spin-down bands) in the (111)-stacking SLs
due to their minimum number of Ni-Ni bonds. By contrast, the spin-down bands
in the (001)-stacking are broader and close to the spin-up channel.

In the $n=1$ case, the narrowing of the spin-down bands
due to quantum confinement gives rise to the observed induced magnetism,
and the effect is clearly the most notorious for the (111)-stacking, as
discussed before.
However, since quantum confinement also exists for the $n=2$ cases,
then why is their induced magnetism so weak even for the (111)-stacking?
Is this nearly-vanishing moment a parametric ``accident'' for the (111) $n=2$ SL
considering its spin-polarized DOS (Fig.~\ref{dos}(f))?
Our analysis suggests that this
nontrivial behavior is caused by the particular quantum properties
of this confined system.
The (111)-stacking perovskite bilayer forms a honeycomb lattice
that has a peculiar band structure.\cite{Xiao:Nc}
In particular, for an isolated LNO bilayer
there is a flat bottom band.\cite{Yang:Prb,Ruegg:Prb,Ruegg:Prb12}
In the current study, although the LNO bilayer is not isolated in the
crystal, its spin-down channel is effectively quantum confined.
Then, the spin-down channel has a nearly flat bottom band
(broaden since the barrier $J_{\rm H}$ is not infinitely high),
that induces a large DOS peak at the band bottom [at $-0.6$ in Fig.~\ref{dos}(f)].
This occupied localized states can accumulate  $0.5$ $e_{\rm g}$ spin-down
electron per Ni which significantly reduces the net induced magnetism.
This tendency is clearly different from the $n=1$ (111)-stacking case,
where the large DOS peak [at $0.8$ in Fig.~\ref{dos}(c)] due to the
confinement is actually unoccupied.

It is also interesting to observe
that this ``accident'' is fairly robust when modifying the parameters in reasonable ranges,
as discussed in the following subsection.
The underlying reason is that here the $e_{\rm g}$ electron density of Ni is higher than $1$,
while the nearly flat bottom band is always far below the Fermi level.
In fact, in the vicinity of the Fermi level, there is no substantial difference
between the spin-up and spin-down DOS.
In summary, the weak induced magnetism for the $n=2$ (111)-stacking
is also due to the quantum confinement.

\begin{figure}
\includegraphics[width=0.5\textwidth]{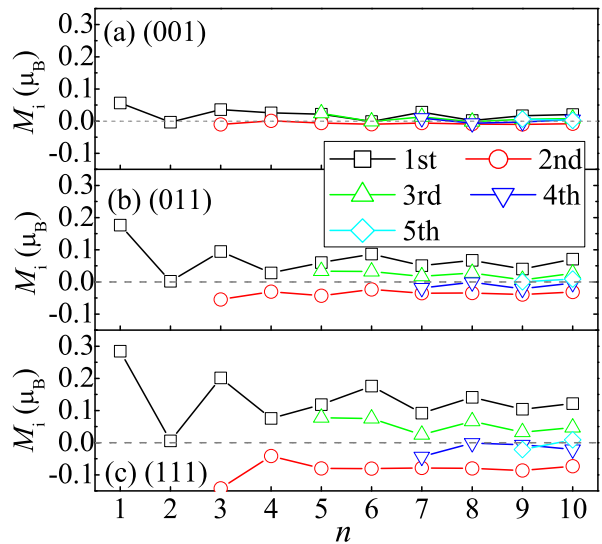}
\caption{(Color online) Magnetization profiles  vs. distances
from the interfaces
(e.g. the ``1st" curves denote the first interfacial layers).
A higher index denotes deeper LNO layers. $V_{\rm Ni}=U=0$ are used.
The stackings are (001) in (a), (011) in (b),
and (111) in (c).}
\label{ml}
\end{figure}

All the results above were obtained for short periodic SLs,
in which all Ni cations are interfacial.
In thicker cases ($n\geqslant3$), there are inner Ni layers that do not connect directly to Mn's.
Then their induced magnetism, if any, can be considered as a second order
effect of the quantum confinement. As shown in Fig.~\ref{ml}, the induced magnetism in these cases
displays an interesting modulation as a function of the distance from the interfaces.
\footnote{For simplify, the long-range Coulombic interaction due to charge transfer is neglected,
which is acceptable if the period $n$ is not too large and the dielectric constant is large.}
As expressed before, the induced magnetization
in the (001)-stacking SLs is always weak,
irrespective of the period $n$.
Therefore, the (001)-stacking LNO/LMO
should not present a robust exchange bias,
in agreement with experiments.\cite{Gibert:Nm} By contrast,
the (111)-stacking SLs display the largest induced
magnetization, with the caveat that it nearly vanishes
at $n=2$ for the reasons already explained.
When $n\geqslant3$, the (111) magnetization of the first layer fluctuates
around $0.1\sim0.2$ $\mu_{\rm B}$ per Ni, while the second Ni layers shows
negative magnetization $\sim -0.1$ $\mu_{\rm B}$ per Ni.
For deeper Ni-layers (3rd, 4th, 5th, and more) the magnetic moments become weaker and weaker,
finally fluctuate around zero and approaching the PM
state of pure LNO. This qualitatively
explains the decreasing exchange bias with increasing period $n$
when $n\geqslant5$ (corresponding to the appearance of the 3rd layers)
observed in experiments.\cite{Gibert:Nm}

Despite the previously described agreement of the interfacial induced magnetism between
our model and the DFT studies,\cite{Gibert:Nm} these two techniques have a
sign difference regarding the induced moments in middle layers of long periodic SLs ($n>5$).
This discrepancy
needs further studies involving both experimental and theoretical components.

\subsection{Correlated effect \& other $V_{\rm Ni}$'s}

\begin{figure}
\includegraphics[width=0.5\textwidth]{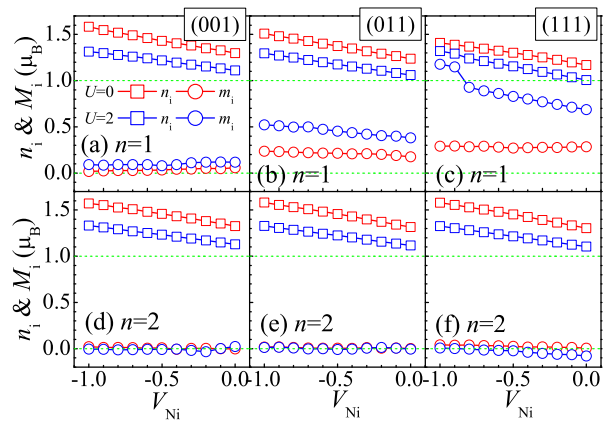}
\caption{(Color online) The Ni $e_{\rm g}$ electron densities (squares) and magnetization (circles)
versus $V_{\rm Ni}$ for $U=0$ (red) and $U=2$ (blue). Top: $n=1$; bottom: $n=2$.
Left/middle/right: (001)/(011)/(111)-stacking.}
\label{uv}
\end{figure}

Our results above were obtained with $V_{\rm Ni}=U=0$. It is necessary
to confirm the robustness of those results with other parameters since
real SLs may correspond to another set of values for $V_{\rm Ni}$ and $U$.
For example, previous DFT studies reported induced moments larger than
those found in our study described above,\cite{Gibert:Nm}
and this may be caused by the influence of correlation effects.

By varying $V_{\rm Ni}$ parametric with $U=2$,
the $e_{\rm g}$ density and $e_{\rm g}$ magnetization of short periodic SLs
are shown in Fig.~\ref{uv}.
Clearly, in all cases, the more negative $V_{\rm Ni}$ becomes, the more $e_{\rm g}$ electrons
accumulate on the Ni layers.
Regarding the induced magnetism, varying $V_{\rm Ni}$ the induced magnetization remains qualitatively robust.
In the $n=1$ cases, the induced moment of Ni is quite weak in the (001) stacking
but prominent in the (111) stacking, as in the $U=0$ case.
With this Hubbard-type correlation effects, the induced moments in
the (111) $n=1$ case are significantly enhanced, result
comparable with the DFT data.

Another effect of the Hubbard interaction is the suppression of the charge
leakage from Mn to Ni. For example, in the $n=1$ case, when $U=2$ and $V_{\rm Ni}=0$,
the $e_{\rm g}$ density of Ni is close to the original value $1$, implying a very weak
charge transfer. However, the induced magnetization remains quite prominent,
further confirming that the charge leakage from Mn to Ni is \emph{not} the key origin
of the induced magnetization in the Ni layers.
In the $n=2$ case, all induced magnetic moments are very weak,
although not exactly zero, even with the Hubbard interaction.
All these results imply that quantum confinement effects are
qualitatively robust within a reasonable parameter region.

\begin{figure}
\includegraphics[width=0.5\textwidth]{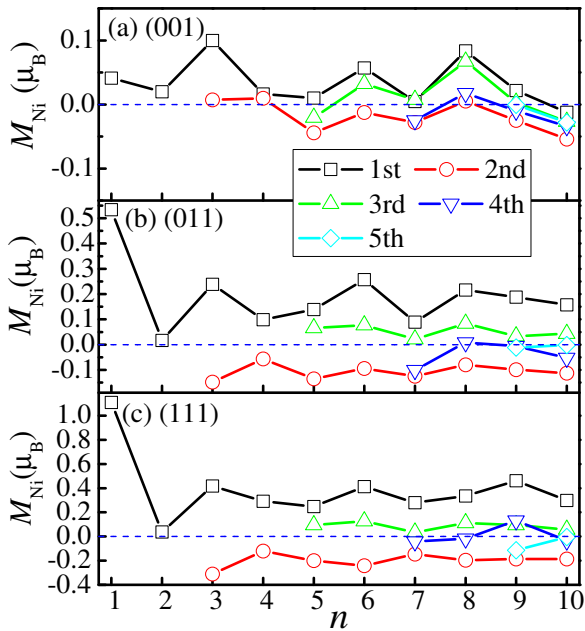}
\caption{(Color online) Magnetization profiles  vs. distances
from the interfaces. $V_{\rm Ni}=-1$, $U=2$ are used.
All notations are the same as in Fig.~\ref{ml}.}
\label{mlu}
\end{figure}

The results for thicker SLs ($n\geqslant3$) show similar behaviors,
with positive induced magnetic moments for the first interfacial layers
followed by weaker negative ones for the second layers.
The thickness-dependent modulation, which is also similar
to the non-correlated limit, is also calculated in the case of thicker SLs
with correlation couplings ($U=2$ and $V_{\rm Ni}=-1$), as shown in Fig.~\ref{mlu}.
Comparing with Fig.~\ref{ml} in the non-correlated limit, no qualitative differences
are observed, although the values of the induced magnetic moments are enhanced
due to the correlation effects. Thus, the oscillatory characteristics
of the induced magnetism with an increasing distance from the interfaces is a
robust feature of our results.

\subsection{RKKY-like exchange}
In Fig.~\ref{ml} and Fig.~\ref{mlu}, two features of the induced magnetization are worth highlighting:
(1) the sign oscillations with increasing
distance from the interfaces and (2) the fluctuations
of the values of the first/second layers with increasing
period $n$, both suggestive of a Ruderman-Kittel-Kasuya-Yosida (RKKY)-like exchange coupling
between the LNO and LMO layers.
In this sense, the almost vanishing magnetism of the LNO bilayer $n=2$
can also be qualitatively understood:
the first Ni layer of the left interface is also the
second layer counting from the right interface,
leading to a partial cancellation of the net magnetization.
This RKKY-based description, which qualitatively
agrees with the previously described explanation based on band structures,
provides a more intuitive understanding than the rather complex calculations
and results presented thus far.

\begin{figure}
\centering
\includegraphics[width=0.5\textwidth]{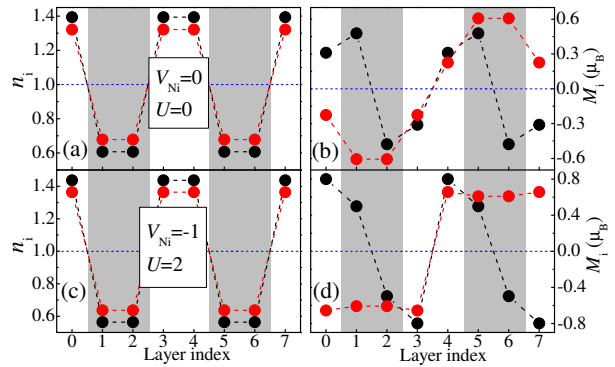}
\caption{(Color online) The profiles of $e_{\rm g}$ density (left)
and $e_{\rm g}$ magnetization (right)
for the $n=2$ (111)-stacking superlattices with antiferromagnetically coupled LaMnO$_3$ layers.
Black: Mn$\uparrow$-Mn$\downarrow$-Ni-Ni-Mn$\uparrow$-Mn$\downarrow$ configuration;
Red: Mn$\downarrow$-Mn$\downarrow$-Ni-Ni-Mn$\uparrow$-Mn$\uparrow$ configuration.
Gray region: LMO;
White region: LNO. (a-b) $V_{\rm Ni}=0$ and $U=0$; (c-d) $V_{\rm Ni}=-1$ and $U=2$.}
\label{n2}
\end{figure}

To confirm this idea, the induced magnetization was recalculated
by flipping the magnetic background
of some Mn layers in the (111)-stacking $n=2$ case.
Two situations were tested: (1) flipping one layer
in each bilayer (Mn$\uparrow$-Mn$\downarrow$-Ni-Ni-Mn$\uparrow$-Mn$\downarrow$);
(2) flipping one bilayer entirely
(Mn$\downarrow$-Mn$\downarrow$-Ni-Ni-Mn$\uparrow$-Mn$\uparrow$). Both these cases now give
very prominent induced local magnetization in
the LNO $n=2$ bilayers, which couples antiferromagnetically between these
two neighboring layers (-Ni$\downarrow$-Ni$\uparrow$-), as shown in Fig.~\ref{n2}. While this result is different from the
ferromagnetically ordered case
(Mn$\uparrow$-Mn$\uparrow$-Ni-Ni-Mn$\uparrow$-Mn$\uparrow$) case,
it also supports the scenario of a RKKY coupling between the Mn's local moments and
the Ni's induced moments.

\begin{figure}
\includegraphics[width=0.5\textwidth]{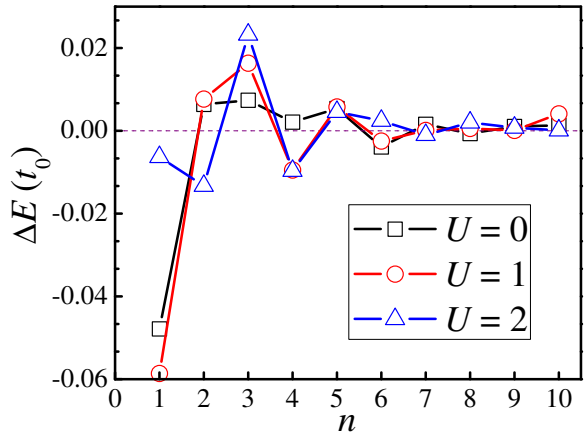}
\caption{(Color online) The energy differences per site between
ferromagnetically coupled and antiferromagnetically coupled
LMO layers ($E_{\rm AFM}-E_{\rm FM}$) in the (111)-stacking superlattices. $V_{\rm Ni}=-1$ is used and $U$ changes from $0$ to $2$.}
\label{eb}
\end{figure}

If the magnetic coupling between LMO and LNO layers is indeed RKKY-like,
then an interesting question arises: how far can the coupling penetrate
in these superlattices?
To address this issue, the energy difference between FM and antiferromagnetic (AFM)
LaMnO$_3$ layers was calculated varying the period $n$.
Here, the AFM coupled LMO layers denote the case
Mn$\downarrow$-...-Mn$\downarrow$-Ni-...-Ni-Mn$\uparrow$-...-Mn$\uparrow$.
As shown in Fig.~\ref{eb}, the absolute energy difference
is large in short-period SLs, but it drops to nearly zero
when $n\geqslant5$. This tendency agrees with the results shown in
Fig.~\ref{ml}(c) and Fig.~\ref{mlu}(c),
where the induced magnetization becomes weak beyond
the 3th layers (corresponding to the $n\geqslant5$ cases).
This result can qualitatively explain the experimentally observed decrease in the exchange
bias with increasing period $n$ for the case $n\geqslant5$.\cite{Gibert:Nm}

\subsection{Additional discussion}

In the computational studies described above, the LMO layers were mainly set
to have a fixed FM spin order, although a few simple AFM cases were also tested.
This  is reasonable considering the experimental evidence that clear FM hysteresis loops
do appear in these SLs.\cite{Gibert:Nm}
However, it is well known that there are various types of AFM phases in doped manganites,
such as A-AFM, CE-AFM, G-AFM, and others, depending
on the doping concentrations and bandwidths.\cite{Hotta:Prb99,Hotta:Prl,Hotta:Prl00,Dong:Prb08,Dong:Prb08.2}
In the LMO-LNO SLs, if the charge leakage is very strong for some particular layers,
it is possible to have any of these types of AFM orders, and considering all these possibilities
is beyond the scope of the present effort.

It is interesting to compare the induced magnetism of the LMO/LNO SLs
against the magnetic reconstruction observed in the LMO/SMO SLs that has been intensively
investigated.\cite{Bhattacharya:Prl,May:Prb,May:Nm,Dong:Prb08.3,Zhang:Prb12,Dong:Prb12}
The induced magnetism in pure manganites SLs
is associated with the magnetic phase transitions
driven by the modification in the $e_{\rm g}$
density and strain.\cite{Dong:Prb08.3,Dong:Prb12,Yu:Prb}
Although not precisely in a one-to-one correspondence,
in pure manganite SLs the induced
magnetism can be traced back to the
phase diagram of bulk manganites.
By contrast, in the LMO/LNO SLs, although experiments
also find a charge transfer from Mn to Ni,\cite{Gibert:Nm,Sanchez:Prb,Hoffman:Arx}
the induced magnetization is nearly decoupled from this $e_{\rm g}$ electron transfer£¬
qualitatively different from the physics scenario in pure manganite SLs.
Instead, here the physics for induced magnetization is more analogous
to the dimensionality control
of electronic phase transitions in LNO-LaAlO$_3$ heterostructures.\cite{Boris:Sci,Liu:Prb}

It is also necessary to mention additional experimental observations.
Despite the experiments of Gibert \textit{et al}, recent related experiments
observed an exchange bias in the (001)-stacking of
La$_{0.75}$Sr$_{0.25}$MnO$_3$/LNO SLs,\cite{Sanchez:Prb} as well as
induced magnetism in (001)-stacked (LMO)$_2$/(LNO)$_n$ SLs,\cite{Hoffman:Arx}
which at least naively seem to disagree with
Gibert \textit{et al}'s experiments and the present simulation.
A probable reason for this discrepancy may reside in the interfacial
intermixing and disorder, which is unavoidable even in the best
state-of-the-art experiments in the oxide interfaces context.
According to our quantum confinement mechanism, at least naively
the disorder may enhance the induced magnetism since more Ni cations
are at interfacial positions in situations of confinement. Of course, additional
experimental and theoretical efforts are necessary in the future to clarify
the role of disorder in this complex system. Certainly incorporating
disorder effects is important but not an easy task in the computational studies
due to the need to repeat the finite cluster calculations dozens of times for
different disorder realizations and average the results to reach
physically relevant results.

\section{Conclusion}

The induced magnetization found in LNO in the LNO/LMO SLs
was studied here via a hybrid microscopic model. The results of our model
agree with previous DFT investigations\cite{Gibert:Nm}
but provide additional details and a deeper physical insight.
Summarizing our conclusions: (1) in the $n=2$ SLs
with FM LMO, the LNO layers are nearly non-magnetic
independently of the stacking directions.
(2) The induced magnetization of LNO in the (111)-stacking SLs
is always the most prominent. By contrast, the LNO layers
in the (001)-stacking SLs are always nearly non-magnetic,
compatible with the exchange bias investigations.
The results for the (011)-stacking are in between those of the (111)-
and (001)-stackings which can be verified in future experiments.
(3) The induced magnetic moments of the first Ni layers
are parallel to the moments of their nearest-neighbor Mn layers, but the
second Ni layers usually display negative moments if not zero.
The induced local magnetism decreases to zero in an oscillatory
manner by increasing
the thickness of the LNO layers. The underlying physical mechanism
for the induced magnetization is associated with
the spin-filtered quantum confinement supplemented by a RKKY-like exchange coupling,
qualitatively different from the magnetization
reconstruction in most previously studied oxide heterostructures.
The present work reported here has emphasized more the approximately parameter-independent
physical results, which may have a broader range of applications to related situations
than the specific study of LMO/LNO may imply.

\begin{acknowledgments}
We thank J.-M. Triscone, P. Zubko, M. Gibert, and A. R\"{u}egg for helpful discussions.
S.D. was supported by the 973 Projects of China (2011CB922101),
NSFC (11004027, 11274060), NCET, and RFDP. E.D. was supported
by the U.S. DOE, Office of Basic Energy Sciences,
Materials Sciences and Engineering Division.
\end{acknowledgments}

\bibliographystyle{apsrev4-1}
\bibliography{../../ref}
\end{document}